\documentstyle[12pt,psfig,amsfonts]{article}
\begin{document}

\topmargin 0pt
\oddsidemargin=-0.4truecm
\evensidemargin=-0.4truecm

\vspace*{-2.0cm}

\vspace*{0.5cm}
\newcommand{\smU}{{\scriptscriptstyle U}}
\newcommand{\sm}[1]{{\scriptscriptstyle #1}}
\def\d{\partial}
\def\l{\left(}
\def\r{\right)}
\newcommand{\e}{\mathop{\rm e}\nolimits}
\newcommand{\Tr}{\mathop{\rm Tr}\nolimits}
\renewcommand{\Im}{\mathop{\rm Im}\nolimits}
\renewcommand{\Re}{\mathop{\rm Re}\nolimits}
\newcommand{\be}{\begin{equation}}
\newcommand{\ee}{\end{equation}}
\newcommand{\ba}{\begin{eqnarray}}
\newcommand{\ea}{\end{eqnarray}}


\begin{center}
  {\Large\bf Orbifold projection in supersymmetric QCD at $N_f\leq N_c$} \\
  \medskip S.~L.~Dubovsky\footnote{E-mail:    \verb|sergd@ms2.inr.ac.ru|}

   \medskip
  {\small
     Institute for Nuclear Research of
         the Russian Academy of Sciences,\\  60th October Anniversary
  Prospect, 7a, 117312 Moscow, Russia
  }
  
\end{center}

\begin{abstract}
Supersymmetric orbifold projection of ${\cal N}=1$ SQCD with
relatively small number of flavors ($N_f\leq N_c$) is considered.  The
purpose is to check whether orbifolding commutes with the infrared
limit. On the one hand, one considers the orbifold projection of SQCD
and obtains the low-energy description of the resulting theory. On the
other hand, one starts with the low-energy effective theory of the
original SQCD, and only then perfoms orbifolding.  It is shown that at
finite $N_c$ the two low-energy theories obtained in these ways are
different.  However, in the case of stabilized run-away vacuum these
two theories are shown to coincide in the large $N_c$ limit.  In the
case of quantum modified moduli space, topological solitons carrying
baryonic charges are present in the orbifolded low-energy theory.
These solitons may restore the correspondence between the two theories
provided that the soliton mass tends to zero in the large $N_c$ limit.
\end{abstract}
{\bf 1.} Recently, a new approach to the dynamics of strongly
coupled gauge theories based on the correspondence between large $N_c$ theories
and supergravity in higher dimensions has been
suggested~\cite{maldacena}. One of impressive results obtained
within this approach is a relation between Green's functions of two
different gauge theories in the large $N_c$ limit. The relation holds
provided the gauge and matter contents are related by
the so called ``orbifold projection''. Then all Green's functions of the
projected (daughter) theory are equal to certain Green's functions of
the original (parent) theory.
Suggested in the framework of string theory
\cite{kachru,vafa1,vafa2,kaku}, this relation has been later supported
by diagrammatic analysis at the level of field theory
\cite{johansen,schmaltz}.  

By making use of supersymmetry breaking orbifold projection, several
non-supersymmetric candidate dual pairs were suggested
\cite{schmaltz} (see \cite{adi} for string theoretical interpretation).  
In order to construct such a pair one starts
with a supersymmetric model exhibiting the Seiberg duality \cite{duality}.
Then one makes orbifold projections of the electric and magnetic
theories and arrives at two apparently different gauge models. The two
models, however, are claimed to be equivalent in the large $N_c$ limit.

A potential caveat of this construction is the necessity to
interchange large $N_c$ and infrared limits. These limits may not
commute if states with masses which scale as $\Lambda/N_c^{\alpha}$, $\alpha>0$,
are present in the spectrum ($\Lambda$ is the infrared scale of the
theory). Such states do not show up in the effective low-energy theory
at finite $N_c$; however, they become massless in the large $N_c$ limit.
This situation is inherent, e.g., in conventional QCD where
$\eta '$-meson becomes massless in the large $N_c$ limit. Another example
is provided by ${\cal N}=2$ supersymmetric theories
\cite{douglas}. 

Yet another
potential problem is that the relation between parent and daughter
theories proven at the level of planar diagrams may be
spoiled by non-perturbative effects.

Several arguments suggesting that these problems do not arise in
${\cal N}=1$ SQCD have been presented in Ref. \cite{schmaltz}.  For
instance, it was shown that large $N_c$ behavior of gluino
condensate and of mesonic Green's functions agrees with the exact results.

Another possible check is to consider the situation where the 
low-energy descriptions of projected original and effective theories are
known. Then one can explicitly check whether these two theories
coincide in the large $N_c$ limit. One example of this type has been
already presented in Ref. \cite{schmaltz}.  Namely, supersymmetry
preserving orbifold projection of SQCD in the region of Seiberg
duality was considered there.  This projection splits both the electric and
magnetic theories into the sets of decoupled theories with smaller numbers of
colors and flavors. These projected theories are again related to each other
by the Seiberg duality, as expected. It is worth noting, that 
the equivalence between the projected theories holds even at finite $N_c$
in this case.

The purpose of this letter is to study whether the equivalence of the
above type holds in ${\cal N}=1$ SQCD with relatively small number of
quark flavors, $N_f\leq N_c$.  In
order to make use of the advantages of supersymmetry and keep
the dynamics under control we restrict our consideration to orbifold
projection that preserves supersymmetry.   
It is shown that at finite $N_c$ the equivalence between the projected
theories does not hold. However, in the case of 
run-away vacuum stabilized by the quark mass term, the two
theories are shown to coincide in the large $N_c$ limit (with
$N_f/N_c$ kept constant).
Consequently,
this  case serves as a
non-trivial check of the commutativity of the large $N_c$ and infrared
limits in ${\cal N}=1$ SQCD. In the case $N_f=N_c$, when quantum
deformation of the classical flat directions occurs, the
orbifolded effective theory does not reproduce the vacuum manifold of
the orbifolded elementary theory at finite $N_c$. 
However, it is shown that topological
solitons exist in this case. If the solitons become massless
in the large $N_c$ limit,  
equivalence between two daughter theories may be restored.
 Finally, it is pointed out that the infrared limit and orbifold
projection do not commute in the special case, when  the parent
theory belongs to the region of the Seiberg duality, while the
daughter theory does not.  This  exception is  due to the
violation of the conditions of the theorem about orbifold projection
in the parent magnetic theory.


{\bf 2.}  Consider ${\cal N}=1$ SQCD with
gauge group $SU(\Gamma N_c)$ and $\Gamma N_f$
flavors of quarks\footnote{In what follows Latin and Greek letters stand for
  flavor and color indices, respectively.} $Q^a_{\alpha}$ and anti-quarks
$\tilde{Q}^{\beta}_b$. $\Gamma$ is
some positive integer. By the large $N_c$ limit we mean the limit
$N_f,\; N_c\to\infty$, with $N_f/N_c$ kept constant. 
A general description of the orbifold
projection can be found in Refs. [2--7].
Let us describe its specific version \cite{schmaltz}
adapted to the case of ${\cal N}=1$ SQCD which we
consider throughout this paper. One makes use of the discrete group ${\mathbb
Z}_{\Gamma}$. The action of the generator of this group on the quark
superfields is defined as follows,
\begin{equation}
\label{action}
Q^a_{\alpha}\to \l T_{N_f}\r_b^a\;
 \l T_{N_c}\r^{\beta *}_{\alpha}Q^b_{\beta}\;,
\end{equation}
where $T_N$ is a $\Gamma N\times \Gamma N$ diagonal matrix which
consists of $\Gamma$ blocks of the size $N\times N$,
\[
T_N= diag(1,\;\e^{i/2\pi\Gamma},\dots,\;\e^{(\Gamma -1)i/2\pi\Gamma})\;.
\]
The extension of this action to the anti-quark and gauge superfields
is straightforward.  The Lagrangian of the orbifolded theory is
obtained from the Lagrangian of the parent theory by removing all
fields and interactions which are not invariant under the 
action\footnote{One can also consider the case when
${\mathbb Z}_{\Gamma}$ is non-trivially embedded into the
non-anomalous $R$-symmetry group. However, supersymmetry is broken in
the daughter theory in the latter case, so we will restrict our
consideration to the case of trivial embedding. } of
${\mathbb Z}_{\Gamma}$. The theorem proven in
Refs. \cite{johansen,schmaltz} says that Green's functions of the ${\mathbb
Z}_{\Gamma}$--invariant fields calculated in the parent and daughter
theories are the same at the level of planar diagrams modulo 
rescaling of coupling constants.
Generally, in order to obtain the relation between parameters of the
parent and daughter theories, one should rescale the fields  in both
theories  in such a way that the corresponding Lagrangians take the form
\begin{equation}
\label{r1}
{\cal L}_p(g_{pi},\;fields)=\Gamma N_cL_p(\bar{g}_{pi},\; fields)
\end{equation}
and 
\begin{equation}
\label{r2}
{\cal L}_o(g_{oi},\;fields)=N_cL_o(\bar{g}_{oi},\;fields)\;,
\end{equation}
where the subscripts $p$ and $o$ are assigned to parent and daughter
(orbifolded) theories.
Then the standard  $N_c$-counting rules imply that couplings $\bar{g}_{pi},\;
\bar{g}_{oi}$ are constant in the large $N_c$ limit. The equivalence
between parent and daughter theories holds provided 
$\bar{g}_{pi}$ and $\bar{g}_{oi}$  are equal.  In SQCD case under
consideration this means that
the canonical gauge coupling constants in the parent  and orbifolded
theories are related as follows,
\begin{equation}
\label{rescaling}
\Gamma g_p^2=g_o^2\;.
\end{equation}

{\bf 3.} Let us start from the case of run-away vacuum. The
parent theory has  $\Gamma N_c$ colors and $\Gamma N_f$ flavors with 
$N_f/N_c<1$. At low energies the dynamics can be described in terms of mesons
\begin{equation}
\label{mesons}
M^a_{b}=Q^a_{\alpha}\tilde{Q}^{\alpha}_b
\end{equation}
with the following effective
superpotential~\cite{ADS-84,NSVZ2} 
\begin{equation} 
\label{mesonsuperpotential} 
W_{eff}=\Gamma\cdot (N_c-N_f)\left(\frac{\Lambda_{h}^{\Gamma(3N_c-N_f)}}{\det M}\right)
^{\frac{1}{\Gamma(N_c-N_f)}}\;, 
\end{equation} 
where $\Lambda_h$ is the holomorphic infrared scale of the theory. The
latter is related to the holomorphic coupling constant $g_h$ in the following
way,
\[
\Lambda_h^{3N_c-N_f}=\mu \e^{-8\pi^2/g_h^2(\mu)}\;,
\]
where $\mu$ is the normalization scale.
The value of the holomorphic coupling constant $g_h$ is determined by 
the Shifman-Vainstein relation \cite{shifman}
between the holomorphic and canonical coupling constants,
\begin{equation}
\label{gc-gh}
\Re\l{8\pi^2\over g_h^2}\r={8\pi^2\over g^2}+N_c\ln g^2\;.
\end{equation}

Besides mesons, the low-energy effective theory contains
 pure Yang-Mills sector corresponding to the
unbroken $SU(\Gamma (N_c-N_f))$ gauge group. 
In order to protect mesons from acquiring infinite vacuum expectation
values, we consider a deformation of the theory by the  mass term 
$mQ^a\bar{Q}_a=m\mbox{Tr}M$ added to the superpotential. We
take this mass term in the flavor-symmetric form  to preserve
the vectorial $SU(\Gamma N_f)$ symmetry relevant to $1/N_c$ expansion. 

Now we apply the orbifold projection to the high-energy theory and
study whether the obtained daughter theory corresponds at low-energies
to the orbifolded effective theory.
Upon orbifolding the high-energy theory one obtains $\Gamma$ decoupled
theories. Each of them has the gauge group $SU(N_c)$ and $N_f$ quark
flavors. 
The rescaling rule (\ref{rescaling}) implies that holomorphic
infrared scale of the orbifolded theory  is equal to
\begin{equation}
\label{lamresc}
\Lambda_o=\Gamma^{-{N_c\over 3N_c-N_f}}\Lambda_h\;.
\end{equation}
The quark mass is left
unchanged under the orbifold projection.

Consequently, at low energies the daughter theory is 
described by the following 
effective superpotential,
\begin{equation}
\label{orbisuperpotential} 
W_{1}=\sum_{i=1}^{\Gamma}(N_c-N_f)
\left(\frac{\Lambda_h^{3N_c-N_f }}{\Gamma^{N_c}\det M^{(i)}}\right)
^{\frac{1}{N_c-N_f}}+\sum_{i=1}^{\Gamma}m\Tr M^{(i)}\;,
\end{equation}
where mesons referring to different gauge sectors are distinguished
by the superscript $(i)$.
In addition there is a pure gluonic sector described by $\left[
SU(N_c-N_f)\right]^{\Gamma}$
gauge group.

Let us now consider the orbifold projection of the effective theory,
the latter
consisting of the mesonic sector described by the superpotential
(\ref{mesonsuperpotential}) and of the gluonic sector described by the
gauge group $SU(\Gamma(N_c-N_f))$. The action of the discrete group 
${\mathbb Z}_{\Gamma}$ on the meson superfields is determined by
Eqs. (\ref{action}) and (\ref{mesons}). 
In the effective meson theory, $\Lambda_h$ serves as the coupling
constant. Consequently, one rescales it according to the general rule
described above. The weak coupling form of the mesonic Kahler
potential, $K(M^{\dagger},M)=2\Tr (M^{\dagger}M)^{1/2}$ \cite{kahler}
(or, equivalently, the explicit relation between the meson and quark superfields
(\ref{mesons})), implies that 
rescaling $M\to \Gamma N_cM$ is needed to cast the Lagrangian of the
effective theory in the form (\ref{r1}). Then, the 
general rescaling rule described above implies that  $\Lambda_h$ again 
rescales according to
Eq. ({\ref{lamresc}).

After the orbifold projection of the low-energy theory one obtains 
the gluonic sector
described by the same $\left[ SU(N_c-N_f)\right]^{\Gamma}$ gauge group as above and
the following meson superpotential,
\begin{equation}
\label{orbieffsuperpotential} 
W_{2}=\Gamma\cdot (N_c-N_f)\prod_{i=1}^{\Gamma}\left(\frac{\Lambda_h^{3N_c-N_f
}}{\Gamma^{N_c}
\det M^{(i)}}\right)
^{\frac{1}{\Gamma(N_c-N_f)}}+\sum_{i=1}^{\Gamma}m\Tr M^{(i)}\;. 
\end{equation}
Obviously, the two superpotentials (\ref{orbisuperpotential}) and
(\ref{orbieffsuperpotential}) are different. However, the
vacuum expectation values of mesons are the same in both cases,
\begin{equation}
\label{vev}
\langle M^{(i)}\rangle=\Gamma^{-1}\Lambda_h^{{3N_c-N_f \over
N_c}}m^{{N_f-N_c\over N_c}}\equiv\langle M\rangle \;.
\end{equation}
Moreover, one can straightforwardly check that the Lagrangians describing
dynamics in these vacua
 are the same for both superpotentials in the large $N_c$ limit. To 
see this, let us compare $F$--terms originating from 
Eqs. (\ref{orbisuperpotential}) and (\ref{orbieffsuperpotential}), which
determine the scalar potentials in the two cases.
Making use of the first superpotential (\ref{orbisuperpotential}) we obtain
\[
F^{(i)}_1\equiv {\d W_1\over\d M^{(i)}}=
-\left(\frac{\Lambda_h^{3N_c-N_f }}{\Gamma^{N_c}\det M^{(i)}}\right)
^{\frac{1}{N_c-N_f}}\l M^{(i)T}\r^{-1}+m\;.
\]
In the case of the second superpotential (\ref{orbieffsuperpotential})
we have
\[
F^{(i)}_2\equiv {\d W_2\over\d M^{(i)}}=-\left(\prod_{j=1}^{\Gamma}\frac{\Lambda_h^{3N_c-N_f
}}{\Gamma^{N_c}
\det M^{(j)}}\right)
^{\frac{1}{\Gamma(N_c-N_f)}}\l M^{(i)T}\r^{-1}+m\;.
\]
We write the meson fields in the form $M^{(i)}=\langle M\rangle+\delta
M^{(i)}$, then $F^{(i)}_1$ and $F^{(i)}_2$  differ by terms coming
from the expansion of determinants in the denominators. However, these
terms are suppressed by extra powers of $N_c-N_f$. 
Hence, both $F$--terms in the large $N_c$ limit are reduced to
\[
F^{(i)}=-\Gamma^{-1}m^{{N_f\over N_c}}\Lambda^{{3N_c-N_f\over
N_c}}\l\langle M\rangle+\delta M^{(i)T}\r^{-1}+m\;.
\]
We conclude that ${\cal N}=1$ SQCD with
$N_f<N_c$ provides a non-trivial check of the technique suggested
in Ref. \cite{schmaltz}.

{\bf 4.} Let us now turn to the case of SQCD with the gauge 
group $SU(\Gamma N_c)$ and $\Gamma N_c$
quark flavors. This theory
exhibits quantum deformation of the
moduli space~\cite{seiberg}. Namely, the space of vacua of the
microscopic theory is described by the set of holomorphic gauge
invariants constructed of quark and anti-quark
fields. These invariants are mesons (\ref{mesons}),
and (anti)-baryons
\[
B=\epsilon^{{\alpha_1..\alpha_
{{\sm{\Gamma N_c}}}}}Q^{1}_{\alpha_1}..Q^{\Gamma N_c}_{\alpha_
{{\sm{\Gamma N_c}}}}\;,
\]
\[
\tilde{B}=\epsilon_{{\beta_1..\beta_
{{\sm{\Gamma N_c}}}}}\tilde{Q}_{1}^{\beta_1}..\tilde{Q}_{\Gamma N_c}^{\beta_
{{\sm{\Gamma N_c}}}}\;,
\]
subject to the constraint
\be
\label{qms}
\det M-B\tilde{B}=\Lambda_h^{2\Gamma N_c}\;.
\ee 
The r.h.s of 
Eq.~(\ref{qms}) is of purely quantum origin and indicates the difference 
between the topologies of the quantum and classical spaces of vacua.
At low energies this theory is described by the
 non-linear sigma model with the 
field space parameterized by mesonic $M^a_{\tilde{b}}$ and (anti-)baryonic
$(\tilde{B})B$ coordinates that satisfy the constraint (\ref{qms}).

Let us take the orbifold projection of this theory with respect to the
discrete group ${\mathbb Z}_{\Gamma}$ as described above. 
Then, in analogy to the case of smaller number of
flavors, the orbifolded theory
splits into $\Gamma$ copies of SQCD with gauge groups $SU(N_c)$ and $N_c$ 
quark flavors. At low energy the latter theory is described 
in terms of sigma-model
fields $M^{a(i)}_{b}$, $B^{(i)},\; \tilde{B}^{(i)}$, where $a,b=1,\dots ,N_c$ 
and $i=1,\dots ,\Gamma$. These fields satisfy the constraints 
\be
\label{qmsi}
\det M^{(i)}-B^{(i)}\tilde{B}^{(i)}=\Lambda_o^{2 N_c}\;
\ee 
for every $i$.

The orbifold projection of the low-energy theory  described by
Eq. (\ref{qms}) is the non-linear
sigma-model with the fields $M^{a(i)}_{b}$, $B,\; \tilde{B}$
subject to the constraint 
\be
\label{qmswrong}
\prod_{i=1}^{\Gamma}\det M^{(i)}-B\tilde{B}=\Lambda_o^{2\Gamma N_c}\;.
\ee 
Manifolds ${\cal Q}_1$ and ${\cal Q}_2$ determined by
Eqs.~(\ref{qmsi}) and (\ref{qmswrong}), respectively, are different. For
instance, the former has complex dimension $\Gamma(N_c^2+1)$ whereas the
complex dimension of ${\cal Q}_2$ is $(\Gamma N_c^2+1)$.  Consequently,
the orbifolded original theory, described at low energies 
by Eq.~(\ref{qmsi}), has extra
$(\Gamma-1)$ massless superfields as compared to the orbifolded sigma
model. This is due to the fact that baryons of the parent sigma model
do not carry flavor or color indices and, as a result, they are
not affected by the orbifold projection. Hence, the
 orbifolded effective theory
has one conserved baryonic charge and one pair of massless baryons
instead of $\Gamma$.

Furthermore, there is a profound difference between ${\cal Q}_1$ and
${\cal Q}_2$. To see this difference, let us study their topology in
more detail.
The manifold ${\cal Q}_1$ is merely a direct product of $\Gamma$ copies of
the manifold ${\cal Q}$ defined by one of the equations listed in
 Eqs.~(\ref{qmsi}). The topology of ${\cal
  Q}$ has been studied in Ref.~\cite{we}. It was found there that this
space is homotopically equivalent to the double suspension of  $SL(N_c,
\mathbb C)$ group, $\Sigma(\Sigma(SL(N_c, \mathbb C)))$. 

The suspension of the manifold $\cal X$ is the cylinder ${\cal
X}\times [0,1]$ with all points on the lower base ${\cal X}\times 0$
identified and all points on the upper base ${\cal X}\times 1$ 
identified as well. For example, the suspension of $d$-dimensional
sphere $S^d$ is $(d+1)$-dimensional sphere $S^{(d+1)}$. The latter
observation is the basis of the theorem about suspension (see., e.g.,
Ref. \cite{fux}, p.79), $\pi_{q+1}\l\Sigma\l{\cal
X}\r\r=\pi_{q}\l{\cal X}\r$ for all $q\leq 2n-2$, provided that
$\pi_i({\cal X}) =0$ for $i<n$. In particular, this theorem implies
that the lowest non-trivial homotopy group of ${\cal Q}$ is
$\pi_5({\cal Q})=\mathbb Z$, so that $\pi_5({\cal Q}_1)={\mathbb
Z}^{\Gamma}$ .

Let us now consider the manifold ${\cal Q}_2$. 
A straightforward generalization of the arguments
of Ref.~\cite{we} shows that this space is homotopically
 equivalent to the double 
suspension of the manifold ${\cal Y}$ determined by the following equation,
\begin{equation}
\label{calX}
\prod_{i=1}^{\Gamma}\det M^{(i)}=\Lambda_o^{2\Gamma N_c}\;.
\end{equation}
At $\Gamma=1$, the manifold $\cal Y$ is $SL(N_c,{\mathbb C})$ in accordance
with the above discussion.

Let us calculate the lowest homotopy group of ${\cal Q}_2$.
 Every non-degenerate matrix $M^{(i)}$ can be
decomposed into a product of a matrix with unit determinant and of
the diagonal matrix of the form $diag(t_i,1,\dots ,1)$. Then
Eq.~(\ref{calX}) implies that the manifold ${\cal Y}$ is
a direct product $SL(N_c,{\mathbb C})^{\Gamma}\times T_{\Gamma-1}$. Here
$T_{\Gamma-1}$ is $(\Gamma-1)$-dimensional complex torus ${\mathbb
C}_*^{\Gamma-1}$ determined by the equation
\[
\prod_{i=1}^{\Gamma}t_i=\Lambda_o^{2\Gamma N_c}\;.
\]
Therefore, the manifold ${\cal Q}_2$ is homotopically equivalent to
\[
{\cal Q}_2\sim\Sigma(\Sigma({\cal Y}))\sim{\cal Q}_1\times (S^3)^{\Gamma-1}\;.
\]
The lowest non-trivial homotopy group of ${\cal Q}_2$ is
\begin{equation}
\label{*}
\pi_3({\cal Q}_2)={\mathbb Z}^{\Gamma-1}\;.
\end{equation}
As a result, contrary to
the case of ${\cal Q}_1$ sigma-model, there may exist topological solitons
in the ${\cal Q}_2$ sigma-model, provided that the stabilizing
higher-derivative terms are present in the Kahler potential of the
effective theory. These solitons are distinguished by $(\Gamma-1)$
conserved topological charges.

It is tempting to identify these charges with the missing $(\Gamma-1)$
baryonic charges. An argument in favor of this identification is
the possibility to construct corresponding topological
currents by formal extension of the algebra of the Noether currents;
in analogy to the case of the conventional QCD \cite{witten}.
This is possible due to the existence of the Wess--Zumino term in SQCD
with $N_f=N_c$. 
In the parent theory this term reads as follows \cite{we},
\[
\Gamma_p={-1\over 12\pi^2\Lambda^{4\Gamma N_c}}\Im\int\!\! d\Omega~
\det\!M\cdot\epsilon^{\mu\nu\lambda\rho\sigma}
\d_{\mu}B\;\;\!\!\d_{\nu}
\tilde{B}\nonumber\times \Tr\l
M^{-1}\partial_{\lambda}M
M^{-1}\partial_{\rho}MM^{-1}\partial_{\sigma}M\r\;. 
\]
After the orbifold projection it takes the following form,
\ba
\Gamma_o={-1\over 12\pi^2\Lambda_o^{4\Gamma N_c}}\Im\int\!\! d\Omega~
\l\prod_{i=1}^{\Gamma}\det\!M^{(i)}\r\cdot\epsilon^{\mu\nu\lambda\rho\sigma}
\d_{\mu}B\;\;\!\!\d_{\nu}
\tilde{B}\nonumber\\\times \sum_{i=1}^{\Gamma}\Tr\l
M^{(i)-1}\partial_{\lambda}M^{(i)}
M^{(i)-1}\partial_{\rho}M^{(i)}M^{(i)-1}\partial_{\sigma}M^{(i)}\r\nonumber\;. 
\ea
The contribution of this term to the
flavor current in the daughter theory reads as follows,
\[
j^{\mu}={1\over 4\pi^2\Lambda_o^{4N_c}}\prod_{i=1}^{\Gamma}\det\! M^{(i)}\cdot
\epsilon^{\mu\nu\lambda\rho} \d_{\nu}B\;\;\!\!\d_{\lambda}\tilde{B}\Tr\l
T^f\sum_{i=1}^{\Gamma}\d_{\rho}M^{(i)}M^{(i)-1}\r+h.c.  \;,
\]
where $T^f$ is a generator of the flavor group.
In complete analogy to the case of QCD, 
one formally substitutes here $T^f$
in the form $diag(0,\dots,i,\dots,0)$ where $0,i$ stand
for the blocks of length $\Gamma$. In this way one obtains $\Gamma$
conserved topological currents,
\[
j^{(i)\mu}={1\over 2\pi^2\Lambda_o^{4N_c}}\Im\prod_{i=1}^{\Gamma}\det\! M^{(i)}\cdot
\epsilon^{\mu\nu\lambda\rho}
\d_{\nu}B\;\;\!\!\d_{\lambda}\tilde{B}\Tr\l
\d_{\rho}M^{(i)}M^{(i)-1}\r \;,
\]
which are subject to the constraint $\sum j^{(i)\mu}=0$. These
currents correspond precisely to the topological property (\ref{*}),
and the above argument indeed shows that they are naturally identified
with the missing baryonic currents.

To prove the equivalence of the ${\cal Q}_1$ and ${\cal Q}_2$
sigma-models in the large $N_c$ limit one would have to study the behavior of
the soliton mass in this limit. The above consideration indicates that
these two models may be  equivalent at large $N_c$ provided that the soliton mass
tends to zero as $N_c\to\infty$. We leave the analysis of this
point for future.

{\bf 5.} Finally, let us present an example in which the orbifold
projection does not commute with the infrared limit. Let us consider
${\cal N}=1$ SQCD with $\Gamma N_c$ colors and $\Gamma (N_c+1)$
flavors. This theory belongs to the region of the Seiberg duality and its 
low-energy behavior can
be described in terms of a magnetic theory with $\Gamma$
colors and $\Gamma (N_c+1)$ flavors. In addition, the  magnetic theory
contains gauge-singlet meson fields $M^a_b$ with the
following superpotential
\[
W_m=qM\tilde{q}\;,
\]
where $q$ and $\tilde{q}$ are dual quark and anti-quark superfields;
a dimensionful coefficient is omitted.

Upon orbifolding the electric theory one obtains $\Gamma$ decoupled theories
with the gauge group $SU(N_c)$ and $(N_c+1)$ quark flavors. At low energies
this theory confines and describes mesons and baryons interacting
through the following superpotential \cite{seiberg}
\begin{equation}
\label{mb}
W_1=\sum_{i=1}^{\Gamma}\l B^{(i)}M^{(i)}\tilde{B}^{(i)}-\det M^{(i)}\r\;.
\end{equation}
The orbifold projection of the magnetic theory, on the other hand,
 leaves $\Gamma$
pairs of the gauge-singlet superfields $q^{(i)}$ and $\tilde{q}^{(i)}$ with
the same quantum numbers as $ B^{(i)}$ and $\tilde{B}^{(i)}$ and
splits the meson multiplet in the same manner as above.
The projected superpotential reads as follows,
\begin{equation}
\label{mq}
W_2=\sum_{i=1}^{\Gamma}q^{(i)}M^{(i)}\tilde{q}^{(i)}\;,
\end{equation}
which is different from Eq.~(\ref{mb}).

The reason of this discrepancy is that 
$1/N_c$ expansion does not work
in the magnetic theory, as the number of magnetic colors does not depend on
$N_c$. Consequently, the relation between the planar diagrams in the parent
and daughter theories does not lead to the relation between 
corresponding Green's functions in this case.

{\bf 6.} To conclude, we have considered orbifold projection of SQCD
with relatively small number of quark flavors.  It was found that in
the case of stabilized run-away vacuum, the orbifold projection serves
as a non-trivial check of the commutativity of the large $N_c$ and
infrared limits.  

We discussed in sect. {\bf 5} also a specific way of taking
$N_c\to\infty$ such that
two daughter theories obtained in different
limits are not equivalent. The reason is that the conditions of the
theorem about the orbifold projection are violated in the effective
theory.
 
The most intriguing situation occurs in SQCD with quantum modified
moduli space, $N_f=N_c$. In this case it was found that upon
orbifolding the high-energy and low-energy theories, one obtains a
pair of sigma models which are not equivalent to each other at finite
$N_c$. Namely, the vacuum manifold of the orbifolded elementary theory
has larger dimension than the space of vacua of the orbifolded
effective theory. As a result, the latter has smaller number of
massless fields in every particular vacuum. Moreover, the number of
the Noether currents in the orbifolded effective theory is too small
to reproduce the correct current algebra of the sigma model
corresponding to orbifolded SQCD.  However, the two sigma models may
become equivalent at large $N_c$ due to the presence of topological
solitons in the orbifolded effective theory. Namely, the 
topological currents restore the structure of the current algebra, and
the solitons may provide the correspondence between the spectra of
light fields in the two theories provided that the soliton mass tends to
zero at large $N_c$. The correct structure of the space of vacua may
be restored due to the presence of topologically non-trivial field
configurations corresponding to non-zero ``soliton condensate''.
Further analysis of this model from both field theoretical and brane
points of view may provide new insights into orbifold projection.

The author acknowledges numerous helpful discussions with F.L.~Bezrukov,
D.S.~Gorbunov, A.A.~Penin and V.A.~Rubakov. This work is supported in
part by Russian Foundation for Basic Research grant 99-02-18410, by
the Russian Academy of Sciences, JRP grant 37 and by ISSEP fellowship.

\def\ijmp#1#2#3{{ Int. Jour. Mod. Phys. }{\bf #1~}(19#2)~#3}
\def\pl#1#2#3{{ Phys. Lett. }{\bf B#1~}(19#2)~#3}
\def\zp#1#2#3{{ Z. Phys. }{\bf C#1~}(19#2)~#3} 
\def\prl#1#2#3{{ Phys. Rev. Lett. }{\bf #1~}(19#2)~#3} 
\def\rmp#1#2#3{{ Rev. Mod. Phys. }{\bf #1~}(19#2)~#3} 
\def\prep#1#2#3{{ Phys. Rep.    }{\bf #1~}(19#2)~#3} 
\def\pr#1#2#3{{ Phys. Rev. }{\bf    D#1~}(19#2)~#3} 
\def\np#1#2#3{{ Nucl. Phys. }{\bf    B#1~}(19#2)~#3} 
\def\mpl#1#2#3{{ Mod. Phys. Lett. }{\bf    #1~}(19#2)~#3} 
\def\arnps#1#2#3{{ Annu. Rev. Nucl. Part. Sci.    }{\bf #1~}(19#2)~#3} 
\def\sjnp#1#2#3{{ Sov. J. Nucl. Phys.    }{\bf #1~}(19#2)~#3} 
\def\jetp#1#2#3{{ JETP Lett. }{\bf    #1~}(19#2)~#3} 
\def\app#1#2#3{{ Acta Phys. Polon. }{\bf    #1~}(19#2)~#3} 
\def\rnc#1#2#3{{ Riv. Nuovo Cim. }{\bf    #1~}(19#2)~#3} 
\def\ap#1#2#3{{ Ann. Phys. }{\bf #1~}(19#2)~#3}
\def\ptp#1#2#3{{ Prog. Theor. Phys. }{\bf #1~}(19#2)~#3}
\def\spu#1#2#3{{ Sov. Phys. Usp.}{\bf #1~}(19#2)~#3}


\begin{thebibliography}{99}
\bibitem{maldacena} J.M.~Maldacena, { Adv.Theor.Math.Phys.}
2 (1998) 231, { Int.J.Theor.Phys.} 38 (1999) 1113
\bibitem{kachru}  S. Kachru, E. Silverstein, \prl{80}{98}{4855} 
\bibitem{vafa1} A. Lawrence, N. Nekrasov, C. Vafa, \np{533}{98}{199}
\bibitem{vafa2}  M.~Bershadsky, Z.~Kakushadze, C.~Vafa,
\np{523}{98}{59}
\bibitem{kaku} Z.~Kakushadze, \np{529}{98}{157}
\bibitem{johansen} M.~Bershadsky, A.~Johansen, \np{536}{98}{141}
\bibitem{schmaltz} M.~Schmaltz, \pr{59}{99}{105018}
\bibitem{adi}  A.~Armoni, B.~Kol, JHEP 9907 (1999) 011
\bibitem{duality} N.~Seiberg, \np{435}{95}{129}
\bibitem{douglas} M.R.~Douglas, S.H.~Shenker, \np{447}{95}{271}
\bibitem{ADS-84} I.~Affleck, M.~Dine, N.~Seiberg, \np{241}{84}{493}
\bibitem{NSVZ2} V.A.~Novikov, M.A.~Shifman, A.I.~Vainshtein and
V.I.~Zakharov, \np{260}{85}{157}
\bibitem{shifman} M.A.~Shifman and A.I.~Vainshtein, \np{277}{86}{456}
\bibitem{kahler}  I.~Affleck, M.~Dine, N.~Seiberg, \np{256}{85}{557}
\bibitem{seiberg} N.~Seiberg, \pr{49}{94}{6857}
\bibitem{we} S.L.~Dubovsky, D.S.~Gorbunov, {\it Induced charge matching and 
Wess-Zumino term on the quantum modified moduli space}, to appear in
{ Phys. Rev.} {\bf D}, hep-th/9909155
\bibitem{fux} A.T.~Fomenko, D.B~Fuchs, {\it A course in homotopy
theory}, Moscow, Nauka, 1989
\bibitem{witten} E.~Witten, \np{223}{83}{422}
\end{thebibliography}
\end{document}